\pdfoutput=1
\documentclass[prd,twocolumn,aps,superscriptaddress,showpacs,preprintnumbers,nofootinbib]{revtex4-1}
\usepackage{latexsym,amsmath,amssymb}
\usepackage{graphicx}
\usepackage[table,usenames,dvipsnames]{xcolor}
\usepackage{hyperref}
\definecolor{azure(colorwheel)}{rgb}{0.0, 0.5, 1.0}
\hypersetup{colorlinks=true,linkcolor=azure(colorwheel),citecolor=azure(colorwheel),pdfborder={0 0 0}}
\usepackage[utf8]{inputenc}
\newcolumntype{P}[1]{>{\centering\arraybackslash}p{#1}}

\makeatletter
\newcommand*{\rom}[1]{\expandafter\@slowromancap\romannumeral #1@}
\makeatother

\newcommand{\ignore}[1]{}

\newcommand{\B}{\mathcal{B}}
\newcommand{\G}{\mathcal{G}}
\newcommand{\Q}{\mathcal{Q}_{\beta_0}}
\newcommand{\Qs}{{\mathcal{Q}_{\beta_\sharp}}}
\newcommand{\Qz}{{\mathcal{Q}_{\beta_{\rm t}}}}
\newcommand{\lcdm}{{\Lambda\mathrm{CDM}}}
\newcommand{\wcdm}{{w\mathrm{CDM}}}
\newcommand{\blcdm}{{\mathrm{B}\Lambda\mathrm{CDM}}}
\newcommand{\cg}{{a_2a_3a_4}}
\newcommand{\axion}{\mathrm{axion}}
\newcommand{\eff}{\mathrm{eff}}
\newcommand{\eg}{{\it e.g.}}

\begin{document}
\title{
  Probing New Physics with High-Redshift Quasars: \\Axions and Non-standard Cosmology
}

\author{Chen Sun}\email{chensun@lanl.gov}
\affiliation{Theoretical Division, Los Alamos National Laboratory, Los Alamos, NM 87545, USA}

\author{Manuel A. Buen-Abad}\email{buenabad@umd.edu}
\affiliation{Maryland Center for Fundamental Physics, Department of Physics, University of Maryland, College Park, MD 20742, U.S.A.}
\affiliation{Dual CP Institute of High Energy Physics, C.P. 28045, Colima, M\'exico}

\author{JiJi Fan}\email{jiji\_fan@brown.edu}
\affiliation{Department of Physics, Brown University, Providence, RI, 02912, USA}
\affiliation{Brown Theoretical Physics Center, Brown University,
Providence, RI, 02912, U.S.A.}




\date{\today}

\begin{abstract}
The Hubble diagram of quasars, as candidates to ``standardizable" candles, has been used to measure the expansion history of the Universe at late times, up to very high redshifts ($z \sim 7$). It has been shown that this history, as inferred from the quasar dataset, deviates at $\gtrsim 3 \sigma$ level from the concordance ($\Lambda$CDM) cosmology model preferred by the cosmic microwave background (CMB) and other datasets. In this article, we investigate whether new physics beyond $\Lambda$CDM ($\blcdm$) or beyond the Standard Model (BSM) could make the quasar data consistent with the concordance model. We first show that an effective redshift-dependent relation between the quasar UV and X-ray luminosities, complementing previous phenomenological work in the literature, can potentially remedy the discrepancy. Such a redshift dependence can be realized in a BSM model with axion-photon conversion in the intergalactic medium (IGM), although the preferred parameter space is {in tension with various other astrophysical constraints on axions, at a level} depending on the specific assumptions made regarding the IGM magnetic field. We briefly discuss a variation of the axion model that could evade these astrophysical constraints. On the other hand, we show that models beyond $\Lambda$CDM such as one with a varying dark energy equation of state ($\wcdm$) or the phenomenological cosmographic model with a polynomial expansion of the luminosity distance, cannot alleviate the tension. The code for our analysis, based on \texttt{emcee}~\cite{Foreman_Mackey_2013} and \texttt{corner.py}~\cite{corner}, is publicly available at \href{https://github.com/ChenSun-Phys/high\_z\_candles.git}{{\tt github.com/ChenSun-Phys/high\_z\_candles}}. 
\end{abstract}


\preprint{LA-UR-23-29579}

\maketitle

\tableofcontents


\section{Introduction}
\label{sec:introduction}

Quasars, or quasi-stellar objects (QSOs), serve as probes in the ultraviolet (UV) and infrared (IR) frequencies up to high redshifts ($z\sim 7$). With some theoretical modeling of their intrinsic luminosities, they can be used to measure luminosity distances as a function of redshift. Recently there has been a renewed interest in using them as ``standardizable" candles to measure the expansion history of the Universe at late times \cite{Risaliti:2018reu,Lusso:2019akb,Yang:2019vgk,Velten:2019vwo,Dainotti:2023ebr,Dainotti:2022rfz,Lenart:2022nip,Bargiacchi:2023jse}. This is of particular importance in light of the recent Hubble tension~\cite{Riess:2016jrr,Riess:2019cxk,Aylor:2018drw,Knox:2019rjx,Abdalla:2022yfra,Krishnan:2020vaf,Colgain:2022nlb}. However, this hope of having a new standard candle has been confronted by a series of challenges in the form of various of consistency checks. In early attempts at its application as a tool to constrain cosmological models, various groups have confirmed that the quasar data prefers an expansion history that stands in stark tension (at $\gtrsim 3 \sigma$ level) with the one
inferred from the cosmic microwave background (CMB)~\cite{Aghanim:2018eyx}, which is in turn consistent with type Ia supernovae (SNIa)~\cite{Pan-STARRS1:2017jku}, and Baryon Acoustic Oscillation (BAO)~\cite{BOSS:2016apd,BOSS:2016hvq} measurements. This discrepancy persists even after the application of stringent cuts to the data that correct for various biases, such as those stemming from dust reddening, X-ray absorption, and Eddington bias~\cite{Lusso:2017hgz}. On the other hand, purely data-driven analyses have been implemented in Refs.~\cite{Dainotti:2022rfz,Lenart:2022nip,Bargiacchi:2023jse}, which correct the luminosity-redshift correlations in the UV and X-ray bands separately. These analyses show that the UV--X-ray relation is indeed robust, and that it does not arise from any luminosity-redshift correlation potentially caused by selection bias. Furthermore they find that, for a given cosmology, the correlation-corrected UV and X-ray luminosities deduced from observations each presents a different redshift evolution, which has been previously unaccounted for. These analyses remain agnostic as to the origin of said evolutions, and limit themselves to characterize their size and impact on the quasar data. In addition, the discrepancy in the quasar data set is also studied in~\cite{Khadka:2020vlh,Khadka:2020whe}. In particular, the redshift where the quasars are standardizable is questioned in \cite{Khadka:2020tlm,Khadka:2021xcc,Khadka:2022aeg,Zajacek:2023qjm}, which shows that the variation of dust extinction plays an important role in the standardization.


In this paper we take a strategy of reverse-engineering: rather than claiming that quasar data favors a cosmology in tension with that from the combined CMB+SNIa+BAO datasets,  we take it as {\it input} in our search for a plausible explanation of the apparent redshift evolution of observed quasar fluxes, in order to restore cosmic concordance. We define the concordance cosmology as the $\lcdm$ model, whose late-time expansion history is fixed by the dark energy density parameter $\Omega_\Lambda$ and the scaling factor $h$ of the Hubble expansion rate $H_0 = 100h \, \mathrm{km/s/Mpc}$~\cite{Aghanim:2018eyx},
\begin{align}
  \label{eq:concordance}
  \Omega_\Lambda = 0.6847 \pm 0.0073~, ~~~ h=0.6736\pm 0.0054~.
\end{align}

While it is possible that the aforementioned luminosity evolution is of a purely astrophysical nature (\eg, coming from a poor understanding of the quasar luminosity itself, a possible temporal bias in quasar formation history, or unaccounted-for propagation effects \cite{Dainotti:2022rfz,Lenart:2022nip,Bargiacchi:2023jse}), in this paper we take a step further and consider whether its origin could be in part coming from new physics. We examine the implications that this evolution has for alternative models beyond $\Lambda$CDM ($\blcdm$) or beyond the Standard Model (SM) of particle physics (BSM). Since the luminosity distance can be affected by both the expansion history of the Universe and photon attenuation, we test three benchmark models: $w$CDM and the cosmographic model \cite{Risaliti:2018reu,Lusso:2019akb} as examples of $\blcdm$, and axion-like particles (ALPs, or axions for short) coupled to photons as an example of BSM~\cite{Csaki:2001yk,Csaki:2001jk,Buen-Abad:2020zbd}. We show that while all these alternatives provide good fits to the QSO dataset, both $w$CDM and the cosmographic model do not resolve the tension between it and the CMB+SNIa+BAO datasets. On the other hand, axion-photon couplings allow for frequency-dependent photon disappearance, which greatly reduces the tension between all the datasets; although it is still in tension with some other astrophysical constraints.




This paper is organized as follows. We briefly review the UV--X-ray relation in quasars and how that is affected by new physics in Sec.~\ref{sec:dist-meas-quas}. We next fit $\lcdm$ to the QSO data with a flexible UV--X-ray relation parametrization in Sec.~\ref{sec:an-effect-evol}, where we reveal the tension between it and other datasets. We demonstrate that, when allowed to change, the quasar data prefers an apparent redshift evolution in the UV--X-ray flux relation, which naturally resolves the aforementioned tension. In Sec.~\ref{sec:bsm-theories} we study the performances of two $\blcdm$ models ($\wcdm$ and cosmographic model) and one BSM model (axion) in alleviating the tension between the QSO and other datasets (SNIa, BAO, CMB). We show that the axion model has the best performance in resolving the tension. We conclude in Sec.~\ref{sec:conclusion}.
The code we used in our numerical analysis, based on \texttt{emcee}~\cite{Foreman_Mackey_2013} and \texttt{corner.py}~\cite{corner}, is publicly available at \href{https://github.com/ChenSun-Phys/high\_z\_candles.git}{{\tt github.com/ChenSun-Phys/high\_z\_candles}}. 

\section{Cosmic Distance Inference with Quasar Data Sets}
\label{sec:dist-meas-quas}

In this section we describe the procedure through which QSO data can be used to determine cosmic luminosity distances as a function of redshift, as well as how new physics can change said procedure.

\subsection{A Lightning Review}
\label{sec:brief-review}

The  observed power-law behavior of the X-ray spectrum of the quasars has motivated the so-called {\it ``two-phase model''}~\cite{Haardt:1991tp,Haardt:1994tx} of the environment surrounding the supermassive black holes (SMBH), which are believed to power the accretion of the active galactic nucleus (AGN). In the two-phase model, ultraviolet (UV) photons are emitted by the relatively cold, optically thick accretion disk around the SMBH; while X-ray photons are produced through up-scattering of the UV photons by the hot, optically thin corona of the SMBH.\footnote{There is a third component from the reflection of the corona photon into the accretion disk peaks around 30~\textrm{keV}\cite{Martocchia:2017bou,Salvestrini:2019thn}.}
  
Since both the UV and X-ray photons are related to the SMBH mass and the accretion rate, their respective luminosities can be formally written as functions of these quantities, namely $L_{\rm X} = f_1(M_{\rm BH}, \dot M_{\rm BH})$, $L_{\rm UV} = f_2(M_{\rm BH}, \dot M_{\rm BH})$~\cite{Lusso:2017hgz}. This implies a relation between $L_{\rm X}$ and $L_{\rm UV}$, as discussed widely in the quasar/AGN literature; see for example Refs.~\cite{zotero-21229,Vignali:2003it,Lusso:2009nq,Steffen:2006px,zotero-21239}. A common parametrization of this relation is the so-called {\it ``Risaliti-Lusso (RL) relation''}, $L_{\rm X} = 10^\beta\, L_{\rm UV}^\gamma$, with $L_{\rm X, UV}$ normalized to $\mathrm{erg/s/Hz}$. The parameter $\gamma$ has been confirmed to have negligible redshift evolution by binning the flux measurement in redshift~\cite{Risaliti:2018reu,Lusso:2017hgz}. In addition, a particular model~\cite{Lusso:2017hgz}, based on the two-phase model~\cite{Haardt:1994tx},
 predicts a redshift-independent $\beta$ as well. 

The RL quasar luminosity relation can also be understood in terms of the \textit{flux} in the UV and X-ray band. 
%
%
%
Under the standard assumption that photon number is conserved, the photon flux $F$ measured by an observer at a luminosity distance $D_L$ can be related to the luminosity $L$ of the photon source as follows:
\begin{align}
  F(z; \omega)
  & = 
    \frac{L(z; \omega)}{4\pi D_L(z)^2} \ ,
\end{align}
where $z$ is the redshift and $\omega$ the photon energy in question.\footnote{We denote the photon flux $F(z; \omega_x)$ at a specific energy $\omega_x$ by $F_x(z)$.}\footnote{Unless otherwise stated, we work in natural units.} 
The RL relation between the quasar UV and X-ray luminosities can then be written in terms of the measured flux as follows:\footnote{We use $\log()$ and $\ln()$ to denote the base-10 and natural logarithms, respectively.}
\begin{align}
  \label{eq:1}
  & \log \left  (\frac{F_{\rm X}(z)}{\mathrm{erg/s/Hz/cm^2}} \right ) -
    \gamma \log \left  (\frac{F_{\rm UV}(z)}{\mathrm{erg/s/Hz/cm^2}} \right )
  \\
  &  =
    2(\gamma - 1) \log \left ({D_L(z)}/{\mathrm{cm}} \right )
    + \beta
    + (\gamma-1) \log(4\pi) \ ,
    \notag
\end{align}
where $\gamma$ and $\beta$, which depend on quasar properties and dynamics, are treated as nuisance parameters. The left hand side depends on the directly observed quasar UV and X-ray fluxes, which means that it can be taken as observational data, modulo the nuisance parameter $\gamma$. The right hand side is a function of $D_L(z)$, which is a prediction of the underlying cosmological model, and of the QSO nuisance parameters $\beta$ and $\gamma$. Thus it can be treated as the theory input.

\subsection{How New Physics Biases the Distance Inference}
\label{sec:how-NP-bias-quasars}

Both astrophysical processes and new physics could alter the {\it flux} relation in Eq.~(\ref{eq:1}) in a redshift-dependent way, which can be understood in terms of an effective $\beta_\eff(z)$. 

If the effective evolution in $\beta_{\rm eff}(z)$ has an astrophysical origin, it must be corrected before the UV--X-ray relation can be reliably utilized for cosmological inferences. This is indeed the subject of Refs.~\cite{Dainotti:2022rfz,Lenart:2022nip,Bargiacchi:2023jse}, where the authors eliminate the luminosity-redshift correlation assuming it takes the form of a given function, and making use of the Efron-Petrosian method~\cite{Efron:1992sch}, before applying the UV--X-ray RL relation. As mentioned in the introduction, these studies showed that the RL relation is not an artifact of any possible selection bias effects, and that there is reason to believe that the UV and X-ray quasar luminosities, as derived from observations and within the context of a specific cosmic history, present each a different, non-negligible redshift evolution. In these investigations, the luminosities in UV and X-ray are assumed to be each corrected by a power-law form,\footnote{More sophisticated functional forms were also tested, which showed no significant difference.} $L_{\rm X, UV} \rightarrow L_{\rm X, UV} (1+z)^{k_{\rm X, UV}}$, to account for the apparent luminosity-redshift evolution. This effectively results in the substitution of the constant $\beta$ with the redshift-dependent $\beta_\eff(z)$:
  \begin{align}
    \label{eq:beta-eff-from-ep-correction}
    \beta \rightarrow \beta_{\rm eff}(z) = \beta + (k_{\rm X} - \gamma k_{\rm UV}) \log(1+z)~. 
\end{align}

However, simply correcting for the luminosity-redshift correlation using data-driven statistical methods, while at the same time remaining agnostic as to its origin, is not suited to our purposes. Indeed, any new physics that may be the source of even just part of this correlation would not be picked up by these methods, but would instead be discarded away during the correction process along with any other less exotic sources. Therefore, we follow a different approach by instead fitting the quasar dataset with different $\beta_{\rm eff}(z)$ functional forms directly predicted by new physics models. 

We consider two classes of new physics models in which this $\beta_\eff(z)$ modification of $\beta$ can arise. The first class consists of deviations from the standard $\Lambda$CDM cosmological expansion history. The measured flux (left-hand side of Eq.~\eqref{eq:1}) constraints the expansion history of the Universe through $D_L(z)$. Anchoring the cosmology to a given concordance history $D_{L,c}(z)$, $\beta_\eff$ can be seen as a deviation from this concordance. Indeed, $\beta$ in Eq.~\eqref{eq:1} is modified to
\begin{align}
  \label{eq:beta-eff-DL}
  \beta \rightarrow \beta_\eff(z) = \beta + 2(\gamma-1) \log[D_L(z)/D_{L,c}(z)] \ .
\end{align}
We would like to point out that such a modification of the expansion history of the Universe would manifest itself in many other independent measurements beyond those of quasars, including CMB anisotropy~\cite{Aghanim:2018eyx}, SNeIa~\cite{Pan-STARRS1:2017jku}, and BAO~\cite{BOSS:2016apd,BOSS:2016hvq,Beutler:2011hx}. As we show below, the quasar data prefers a non-standard cosmology at odds with that favored by these other observations, which effectively limits this interpretation of the tension.

The second class of models involves frequency-dependent propagation effects that result in photon disappearance, which can also break the {\it flux} relation in Eq.~\eqref{eq:1} without violating the relation in the {\it luminosities} $L_{\rm X}$ and $L_{\rm UV}$:
\begin{align}
  \label{eq:extra-attenuation}
  F(z; \omega)
  & = P_{\gamma\gamma}(z; \omega)
    \frac{L(z; \omega)}{4\pi D_L(z)^2} \ ,
\end{align}
where $P_{\gamma\gamma}(z; \omega)$ is the photon survival probability. This effectively introduces a redshift dependence in $\beta$ as follows:
\begin{align}
  \label{eq:beta-eff-axion}
\beta \rightarrow \beta_\eff(z) = \beta + \log \, P_{\rm \gamma\gamma}(z; \omega_{\rm X}) - \gamma \log \,P_{\rm \gamma\gamma}(z, \omega_{\rm UV}) \ .
\end{align}
%
In other words, if there is unaccounted-for extra attenuation in the measured photon flux, a preference for non-zero en-route photon disappearance will arise within the context of a given concordance cosmological expansion history, such as $\Lambda$CDM; and it will manifest itself as a redshift dependence of $\beta$ in the quasar data. In the scenario where this photon disappearance is caused by new particles, the observed flux's dependence on $P_{\gamma\gamma}(z; \omega)$ effectively turns quasars into a probe of BSM physics.

In this paper we take axion-photon interactions as a benchmark scenario of BSM physics responsible for photon disappearance, and test for axion-photon conversion in inter-galactic medium (IGM) using the QSO dataset. We will see how this model, in conjunction with the standard $\Lambda$CDM, provides a very good fit to the data. While this fit is in tension with some astrophysical constraints on the axion-photon coupling, axion-photon interactions provide a concrete physics model that both greatly improves the fit to the QSO data and restores the consistency between the expansion history preferred by this data and that favored by CMB, SNIa, and BAO. We also compare its goodness of fit with a few $\blcdm$ models.

To summarize, our approach differs from and complements the analyses in Refs.~\cite{Dainotti:2022rfz,Lenart:2022nip,Bargiacchi:2023jse} in a few ways. Firstly, Refs.~\cite{Dainotti:2022rfz,Lenart:2022nip,Bargiacchi:2023jse} designed a data-driven method to empirically correct any luminosity-redshift corrections. This results in the elimination of both any selection bias {\it and} any new physics possibly contributing to this redshift evolution. Thus, this method could potentially hide any new physics--induced flux evolution, should there be any. 
Secondly, it is not our purpose to understand all luminosity-redshift correlation present in the quasar data. By fitting the data with new physics-induced $\beta_{\rm eff}$, we perform a critical examination of whether \textit{part} of the flux-redshift evolution may be caused by new physics, if the concordance cosmological history encoded in $\Lambda$CDM and Eq.~(\ref{eq:concordance}) indeed describe nature.
Lastly, while Refs.~\cite{Dainotti:2022rfz,Lenart:2022nip,Bargiacchi:2023jse} are of critical importance in establishing the UV--X-ray RL relation on firmer ground, the templates of the luminosity-redshift evolution adopted therein only allow for smooth apparent luminosity changes across a relatively large range of redshifts. As will become clear later in this paper, the new physics model we test cannot be captured by the simpler functional forms tested in Ref.~\cite{Dainotti:2022rfz}. In particular, the axion-induced flux attenuation generates a somewhat abrupt change in both the UV and the X-ray bands, more akin to a step function than to a slowly-varying transition.

\section{An Effective Evolution of the UV--X-ray Relation}
\label{sec:an-effect-evol}

In this section, we first discuss the tension between QSO dataset and CMB+SNIa+BAO within the context of $\Lambda$CDM. We then show how an effective $\beta_\eff(z)$ with redshift dependence is preferred by quasars, once the cosmological expansion history is itself anchored by CMB+SNIa+BAO. The datasets we use include:
\begin{itemize}
\item $\B$: Our baseline datasets. These include SNIa: Pantheon~\cite{Pan-STARRS1:2017jku}; BAO: 6dFGS~\cite{Beutler:2011hx}, SDSS using the MGS galaxy sample~\cite{Ross:2014qpa}, CMASS and LOWZ galaxy samples of SDSS-III DR12 \cite{Alam:2016hwk}.

\item $\G$: Gaussian priors including $H_0 = 67.36 \pm 0.54 \,\mathrm{km/s/Mpc}$ and $r_s=147.09 \pm 0.26 \,\mathrm{Mpc}$ from Planck 2018~\cite{Aghanim:2018eyx}.\footnote{We also tested the Gaussian prior of $H_0$ from SH0ES, which is in significant tension with the Planck results. We find the same results. This is expected as only the shape of the Hubble diagram matters in our analysis. }

\item $\mathcal{Q}$: QSO dataset from Ref.~\cite{Lusso:2020pdb}. We further parametrize the nuisance $\beta$ parameter in three possible ways:
  \begin{itemize}
  \item $\Q$: $\beta_\eff = \beta_0$ is a constant across all redshifts. This is the most common parametrization in the quasar literature.
  \item $\Qs$: $\beta_\eff$ is a step function. At redshift $z_0$, it sharply transitions from a value $\beta_0$ to $\beta_1$. 
  \item $\Qz$: $\beta_\eff$ goes through a smooth transition, parametrized with a $\tanh$ function as follows
    \begin{align}
      ~~~~~~~~~~ \beta_\eff(z) = \beta_0 + \frac{\beta_0-\beta_1}{2} \left [  \tanh \left ( \frac{z-z_0}{\delta z} \right )  +1 \right ].
    \end{align}
   According to this parametrization there are five nuisance parameters in the QSO dataset in total: $\gamma$, $\beta_0$, $\beta_1$, $z_0$, and $\delta z$.
  \end{itemize}
\end{itemize}


The details of the fits can be found in Appendices~\ref{sec:quasar-data-set} and \ref{sec:methodology-1}. The 1D posteriors can be found in Appendix~\ref{sec:corner-plots}. We list the log-likelihood of the best fit points and the posteriors of dark energy density parameter, $\Omega_\Lambda$, in seven different runs with $\Lambda$CDM in Tab.~\ref{tab:lcdm-fits}. 
%
\begin{table}[ht]
  \centering
  \begin{tabular}{c|P{1.5cm} P{1.5cm} P{1.5cm}|P{1.5cm}}
    \hline 
    & QSO & SNIa & BAO  & $\Omega_\Lambda$\\
    \hline
    $\B+\G$ &  & -1177.4 & -9.2  & $0.68 \pm 0.02$\\    
    $\Q+\G$   & -106.6 & &  & $0.05 ^{+0.07}_{-0.03}$ \\
    $\Qs+\G$  & -169.2 & &  & $0.10^{+0.12}_{-0.07}$ \\
    $\Qz+\G$   &  -179.1 & &  & $0.46^{+0.26}_{-0.27}$ \\
    $\Q+\B+\G$ & -66.62 & -1174.9 & -9.4 & $0.66\pm 0.02$  \\
    $\Qs+\B+\G$ & -149.2 & -1177.0 & -9.4 & $ 0.67\pm 0.02$\\
    $\Qz+\B+\G$ & -179.1 & -1177.3 &  -8.9 & $0.68 \pm 0.02 $ \\
    \hline 
  \end{tabular}
  \caption{\label{tab:lcdm-fits}
The log-likelihood of the best fit points in seven different runs with $\Lambda$CDM. 
}
\end{table}
%


We make a few comments on the results below. First, the aforementioned tension between the QSO and SNIa+BAO datasets can be easily seen in the posterior of $\Omega_\Lambda$ from the $\Lambda$CDM fit to $\B+\G$ ($\Omega_\Lambda = 0.68 \pm 0.02$) and to $\Q+\G$ ($\Omega_\Lambda= 0.05 ^{+0.07}_{-0.03}$). Although $\Q+\B+\G$ leads to a posterior in $\Omega_\Lambda$ that is consistent with the concordance value, this is mostly due to the addition of SNIa+BAO. This can be seen by comparing the log-likelihood between $\Q+\B+\G$ and $\Q+\G$. The fit to quasars is degraded significantly once the baseline data combination SNIa and BAO is added, with a change of $\Delta \chi^2 = + 40.0$ for the QSO dataset alone. What is more, the fit to SNIa within the $\Q+\B+\G$ run is also slightly worse than that within $\B+\G$, $\Delta \chi^2 = +2.5$. This is due to quasars pulling the parameters away from the minimum of SNIa. Put together, these facts are indications of the incompatibility between the QSO dataset assuming a constant $\beta$ and SNIa+BAO. 

Second, the tension is greatly relieved when considering a redshift-dependent $\beta_\eff(z)$: 
\begin{itemize}
\item In the $\Qs+\G$ run, there is a clear preferred transition point at $z_0 = 1.65^{+0.01}_{-0.01}$. While the posterior in $\Omega_\Lambda$ is still largely off compared with that from $\B+\G$, there is huge improvement in the fit  to the QSO dataset with $\Delta \chi^2 = -62.6$, at the cost of only two more parameters. Furthermore, by comparing $\Qs+\B+\G$ and $\Qs+\G$, we see the former has only a milder deterioration in the log-likelihood of quasars, $\Delta \chi^2 = 20.0$, while the value of $\Omega_\Lambda$ is consistent with that from $\B+\G$. 
\item With $\Qz$, the tension is completely resolved. Comparing $\Qz+\G$ with $\Q+\G$, we see an improvement in the fit to the quasar data with $\Delta \chi^2 = -72.5$ between the two fits. In addition, the fit to quasars is as good in $\Qz+\B+\G$ as in $\Qz+\G$. More importantly, $\Qz+\G$ leads to a value of $\Omega_\Lambda$ that is compatible with the concordance cosmology. We show the corresponding shape of $\beta_\eff(z)$ in Sec.~\ref{sec:effect-evol-beta}. 
 \end{itemize}

\section{Implications for B$\Lambda$CDM and BSM Models}
\label{sec:bsm-theories}

In this section, we discuss several specific $\blcdm$ and BSM models that realize an effective redshift-dependent $\beta_\eff(z)$, and whether they could alleviate the tension between the QSO dataset and other cosmological datasets. 

\subsection{Beyond $\Lambda$CDM}
\label{sec:wcdm}

The luminosity distance can be computed once a cosmological expansion history is given:
\begin{align}
  \label{eq:lum-distance}
  D_L(z)
  & =
    (1+z) \, \int_0^z \, dz' \frac{c}{H(z')}~,
\end{align}
where $c$ is the speed of light. This affects the UV--X-ray relation in the measured fluxes as shown in Eq.~\eqref{eq:1}, which causes an effective modification of the QSO nuisance parameter $\beta$ given in Eq.~\eqref{eq:beta-eff-DL}.

First we consider the case of $w$CDM, where the equation of state of the dark energy, $w$, deviates from $-1$. 
A phenomenological parametrization commonly used in the literature is
\begin{align}
  \label{eq:w}
  w(a)
  & = w_0 + w_a ( 1-a)~,
\end{align}
where $a$ is the scale factor. This function smoothly interpolates the current equation of state (EOS) $w_0$ to its value in the early Universe $(w_0+w_a)$. The parameters $w_0, w_a$ are allowed to vary between -1 and +1. The $w$CDM model has therefore four theory parameters to be fitted, $\Omega_\Lambda, H_0, w_0, w_a$.

We also follow Refs.~\cite{Risaliti:2018reu,Lusso:2019akb} and perform a free-form polynomial expansion of the luminosity distance in order to study potential alternative cosmologies in a more general way. If one fixes the low redshift luminosity distance to $c/H_0$, the so-called cosmographic approach is an effective expansion of the form:
%
\begin{align}
  D_L(z)
  & =
    \frac{c}{H_0}
    \bigg ( \ln(1+z) + \sum_{i=2} a_i \ln^i(1+z) \bigg )~. 
\end{align}
We truncate the expansion up to the forth order. Therefore, the model parameters are $H_0, a_2, a_3, a_4$. Note that it is known that the convergence of the series is poor when they are mapped to physical models ($\Lambda$CDM, $w$CDM, etc.)~\cite{Cattoen:2007sk,Yang:2019vgk,Banerjee:2020bjq,OColgain:2021pyh}. Therefore, we use this only as a benchmark phenomenological parametrization to compare with the literature, and refrain from making any statements about the underlying cosmological model.

\subsection{Beyond SM}
\label{sec:beyond-sm}

We now devote our attention to a concrete particle physics model that can leave imprints in the UV--X-ray relation of Eq.~\eqref{eq:1}. If axions exist and couple to photons, the number of photons may not be conserved when they propagate through a static magnetic field background, such as the IGM magnetic field. 
The photon disappearance probability $P_0$ (within a single magnetic domain) due to an axion $a$ with a mass $m_a$ and a coupling to photons $g_{a\gamma} a F \tilde{F}/4$, where $g_{a\gamma}$ is the constant coupling with energy dimension $-1$, and $F$ ($\tilde{F}$) is the (dual) electromagnetic field strength, is given by the well-known formula \cite{Georgi:1983sy,Sikivie:1983ip,Raffelt:1987im,Csaki:2001yk}:
%
\begin{align}
  \label{eq:pag}
    P_0 = \frac{(2\Delta_{a\gamma})^2}{k^2} \sin^2 \left  ( \frac{k x}{2} \right ) \ ,
\end{align}
where $x$ is the distance traveled by the photon, and
\begin{align}
\label{eq:oscs}
    k & \equiv \sqrt{ (2\Delta_{a\gamma})^2 + \left  ( \Delta_a - \Delta_\gamma \right )^2 } \ , \\
    \Delta_{a\gamma} & \equiv  \frac{g_{a\gamma} B}{2} \ , \quad \Delta_a \equiv \frac{m_a^2}{2 \omega} \ , \quad \Delta_\gamma \equiv \frac{m_\gamma^2}{2 \omega} \ ,  
\end{align}
with $B$ the IGM magnetic field transverse to the photon trajectory, and $\omega$ the photon energy. Here $m_\gamma^2 \equiv \frac{4 \pi \alpha n_e}{m_e}$, where $m_e$ and $\alpha$ are the electron mass and fine-structure constant respectively, is the effective photon mass squared in the presence of an ionized plasma with an electron number density $n_e$. In our analysis, we take $n_e \simeq 1.6\times 10^{-8}\,\mathrm{cm}^{-3}$~\cite{Nicastro:2018eam,Martizzi:2018iik}. There are therefore a total of four theory parameters to be fitted $\Omega_\Lambda, H_0, m_a, g_{a\gamma}$.

We adopt the \textit{cell model} for the IGM magnetic field, described in Refs.~\cite{Csaki:2001yk,Csaki:2001jk,Grossman:2002by,Buen-Abad:2020zbd}. In this model the magnetic field is assumed to be split into domains (``cells''), in which it can be taken to be homogeneous. The photon path, extending from a source at some distance $y$ to the observer, is assumed to cross a large number $N$ of these magnetic domains. Each {\it i}-th domain has a {\it physical} size $L_i$ and a randomly oriented magnetic field of strength $B_i$~\cite{Grossman:2002by}, whose component perpendicular to the photon's path is assumed to be the same in each domain. With these simplifications, the resulting net probability of photon-axion conversion over many domains is then given by \cite{Avgoustidis:2010ju}
\begin{align}
  \label{eq:pconv_prod}
  P_{a\gamma}(y) = (1-A) \left  (  1 - \prod\limits_{i=1}^{N} \left  ( 1 - \frac{3}{2}P_{0,i} \right ) \right ) \ ,
\end{align}
where $A \equiv \frac{2}{3} \big ( 1 + \frac{I_a^0}{I_\gamma^0} \big )$ depends on the ratio of the initial intensities of axions and photons coming from the source, denoted by $I_a^0$ and $I_\gamma^0$ respectively; and $P_{0,i}$ is the conversion probability in the {\it i}-th magnetic domain, which can be obtained from Eq.~\eqref{eq:pag} for $x = L_i$. 
Since $N$ is very large, Eq.~\eqref{eq:pconv_prod} can be rewritten as an integral. In order to do this, we further assume that $y$ is a distance that scales linearly with $N$, such that $y/N$ remains constant as $N$ goes to infinity. For example, for IGM propagation the domains are typically assumed to be evenly distributed in {\it comoving} space, which means that each domain has comoving size $s$ and the distance to the source is a comoving distance $y=Ns$. Under these assumptions, we have
\begin{align}
  \label{eq:pag_2}
  P_{a\gamma}(y) = (1-A) \left  ( 1 - \exp \left  [ \frac{1}{s} \int\limits_0^y \mathrm{d} y' ~ \ln \left  ( 1 - \frac{3}{2} P_0(y') \right ) \right ] \right ) \ .
\end{align}
We assume $A=2/3$ or equivalently $I_a^0=0$ throughout the paper.
The ratio of the observed photon flux and the emitted photon flux from the source is then given by
$  P_{\gamma\gamma} = 1 - P_{a\gamma}$.

Let us now briefly comment on the coherent length of the magnetic field domain. While the existence of IGM magnetic fields has been indirectly attested (from the absence of $\gamma$-ray cascade emission) and there are upper bounds as well (from CMB anisotropy as well as turbulence decay), observational constraints on the coherent length are lacking.
{We now briefly discuss a few observable consequences of a sizable IGM magnetic field with a domain size above Mpc. See for example~\cite{Durrer:2013pga} for a more thorough review. Large magnetic domains lead to large Faraday rotations of polarized radio emission from distant quasars, which scales as $(s \mathcal{H})^{-1/2}$ where $\mathcal{H}$ is the comoving Hubble constant. The two leading challenges in utilizing distant quasars to measure the magnetic field in IGM are the determination of line-of-sight electron number density and subtraction of the effect due to the galactic magnetic field. By taking into account the information obtained from the Lyman-$\alpha$ forest, the magnetic field is limited to be below $\mathcal{O}(\mathrm{nG})$ with a domain size smaller than the horizon size $1/H_0$. Next generation radio telescopes are likely to improve our understanding of the galactic magnetic field's contribution to the rotation measure, which can lead to a better extraction of the IGM contribution. In addition, CMB observables (anisotropy, spectral distortion, polarization, and Sunyaev-Zeldovich effect) can be combined to give a comparable limit on the IGM magnetic field for the domain sizes ranging between Mpc and Gpc. Another effect of the IGM magnetic field is on ultra high energy cosmic rays (UHECR), whose trajectory can be deflected by the former. A magnetic field with a coherent length larger than their propagation length can lead to significant deflection of the UHECR. When the coherent length is smaller than the distance of the UHECR source, the cosmic ray experiences random deflections which leads to the deflection angle suppression $\propto \sqrt{s}$. Future UHECR telescopes are likely to further improve the constraints of the IGM magnetic field with a domain size between Mpc to Gpc by another 1-2 orders of magnitude.}
From the theoretical perspective, {magnetic hydrodynamic simulations show that large power of magnetic field at small scales $\ll \mathrm{Mpc}$ drives turbulence in the primordial plasma and IGM, which removes power at small distance scales and increase power at large scales. It is estimated to be below $10\,$nG at $s=1\,\mathrm{Mpc}$ and the limit relaxes quickly as $\propto s$ at larger scales. Since} the coherent length of a magnetic field is related to the structures that support the magnetic field, it is reasonable to believe that the magnetic fields at the intersections of filaments has a domain size of $\sim$Mpc scale. Nevertheless, the IGM magnetic field domains can be much larger due to characteristic size of other structures in the IGM. Because voids and sheets make up the majority of the IGM volume, it is natural to suppose that the relevant length scale associated with their corresponding magnetic fields should be related to the size of these structures. From the IllustrisTNG simulation\cite{Martizzi:2018iik}, one can see that the voids and sheets at low redshift ($z\sim 1$) can reach sizes of $\mathcal{O}(10)$~Mpc. In Sec.~\ref{sec:more-axion-model}, we fit the axion model to our data sets {with domain sizes $s \in \{ 1, 10\}\,\mathrm{Mpc}$ and $B=1\,\mathrm{nG}$.}
Elsewhere, and unless stated otherwise, we use as a benchmark a magnetic domain size of $1~\mathrm{Mpc}$ and a magnitude field of $B = 1\,\mathrm{nG}$ throughout this paper. 

At last, we stress that the impact of axion-photon interactions only manifests itself in distance measurements that involve photon flux, such as the luminosity distances of quasars. It is otherwise largely invisible in experiments based on direct measurements of the angular diameter sizes (\textit{e.g.} CMB observations.)
In addition, for the axion-photon couplings in which we are interested, the impact on SNIa luminosity distances (sensitive to only $\mathrm{eV}$-energy photons) is negligible, most effects being limited to the UV photons at $z>1.5$.




\subsection{Evaluation of the Fits}


We fit each model ($w$CDM, cosmographic, and axion models) to either quasars alone $(\Q+\G)$ or to quasars and our baseline datasets ($\Q+\B+\G$), including in both cases the Gaussian prior ($\G$) on $H_0$ and $r_s$. See Appendices \ref{sec:quasar-data-set} and \ref{sec:methodology-1} for details on our methodology, including the QSO dataset, the priors, and the fitting procedure.
We summarize the resulting log-likelihoods of the best fit points in Tab.~\ref{tab:bcdm-bsm-fits}. The 1D and 2D posteriors are shown in Appendix~\ref{sec:corner-plots}.
\label{sec:results-blcdm-bsm}
\begin{table}[ht]
  \centering
  \begin{tabular}{l|P{1.2cm} P{1.1cm} P{1.1cm}|P{1.5cm}}
    \hline 
    & QSO & SNIa & BAO  & $ \chi^2_c$\\
    \hline
    \hline
    $\wcdm(\Q+\G)$ & -171.1  &  &   &  1247.5\\
    $\wcdm(\Q+\B+\G)$ & -91.7 & -1172.0  & -9.1  & 743.0\\
    \hline
    $\cg(\Q+\G)$ & -179.4  & &   & 996.9 \\
    $\cg(\Q+\B+\G)$ & -140.5  & -1167.7 & -8.9  & 822.5\\
    \hline
    \hline
    $\axion(\Q+\G)$ & -169.7  &  &   &  609.0\\
    $\axion(\Q+\B+\G)$ & -147.8  & -1176.5  & -9.3  & 514.2\\
    \hline 
  \end{tabular}
  \caption{\label{tab:bcdm-bsm-fits}
    The log-likelihood of the best fit points in six tests with beyond $\Lambda$CDM and beyond SM models. The cosmographic model is labeled as $\cg$. In the last column, we show the ``theory distance'' between the best fit points and the concordance cosmological model by comparing their $\chi^2$ difference from the quasars likelihood; see Eq.~(\ref{eq:chi2c}). The smaller the value is, the closer the preferred cosmology is to the concordance model. 
  }
\end{table}



\textbf{Tension between QSO and SNIa+BAO.}~~~~ We start by noting that the axion model significantly reduces the tension between QSOs ($\Q$) and SNIa+BAO ({$\B$}). In comparison, the stronger tension between $\Q$ and $\B$ remains in $w$CDM and the cosmographic model. 

This is reflected in the goodness of fit to $\Q$ of each model. When SNIa and BAO are included, the goodness of fit to QSOs is degraded by $\Delta \chi^2 = 79.4$ for $w$CDM, by $\Delta \chi^2 = 38.9$ for the cosmographic model, and by $\Delta \chi^2 = 21.9$ for the axion model. This shows that both $w$CDM and the cosmographic model take a significant deviation from the concordance cosmology over a large range of redshifts in order to accommodate the QSO dataset, with such changes disfavored by SNIa+BAO. As a result, including SNIa+BAO limits how close these models can reach the minimum of the quasar likelihood.

On the other hand, the axion model allows a modification of the UV and optical photon flux starting at $z\gtrsim 1.5$ while leaving the low-$z$ part mostly unchanged. This minimizes the impact on SNIa, and BAO is not affected by it at all. As a result, including SNIa+BAO only has a mild effect on the goodness of fit to QSO data. Such ``flexibility'' of the axion model compared to $\wcdm$ and the cosmographic model is the key to reduce the tension between the $\Q$ and $\B$ datasets. Meanwhile, the flexibility of the axion model comes at no extra cost in terms of the introduction of more parameters than the $\wcdm$ and cosmographic models.

It is interesting to note that the fit to quasar in $\cg(\Q+\G)$ and $\axion(\Q+G)$ only differ by less than $\Delta \chi^2 = 10$, with the $\cg$ model performing slightly better. However, when $\B$ is included, the axion model is preferred with $\Delta \chi^2 \approx -7.3$, due to precisely the aforementioned flexibility in the axion model that effectively decouples the impact on quasar flux, SNIa flux, and BAO.

\textbf{Restoration of the Concordance Cosmology.}~~~~
Resolving the tension means the minimum of $\Q$ should get closer to that of $\B$ in the theory space. However, this does not guarantee a restoration of concordance cosmology. For example, the minima of the two datasets can both drift toward a theory point that is far away from the concordance model. 
Therefore, aside from the relative distance between the minima of $\Q$ and $\B$, we would also like to evaluate the distance between the minima and a fixed point anchored by the concordance model, which is essentially the minimum of the Planck likelihood. 

To quantify their agreement to the well-established concordance cosmology given by Eq.~\eqref{eq:concordance}, we compare the distance modulus computed from the best fit points of the six runs $\{\cg, \wcdm, \axion\} \otimes \{\Q+\G, \Q+\B+\G\}$ with that computed from Eq.~\eqref{eq:concordance}. More concretely, we define the concordance distance modulus as
\begin{align}
  \label{eq:concordance-dist-modulus}
    \mu_c
  & = 5\; \log_{10} \left (\frac{D_{L,c}}{10\,\mathrm{pc}} \right ), 
\end{align}
where $D_{L,c}$ is the luminosity distance computed from Eq.~\eqref{eq:lum-distance} with the Hubble constant taken from Eq.~\eqref{eq:concordance}. The error associated with $\mu_c$ is given by
\begin{align}
  \label{eq:error-mu-c}
  \Delta \mu_c = \left [ \left ( \frac{\partial \mu_c}{\partial \Omega_\Lambda} \right )^2 \left(\Delta \Omega_\Lambda\right)^2
  + \left ( \frac{\partial \mu_c}{\partial h} \right )^2 \left(\Delta h\right)^2
  \right ]^{1/2}.
\end{align}
The distance modulus at each best fit point of the six runs can be computed with the $D_L$ inferred from the cosmological parameters [$(a_2, a_3, a_4)$ in the cosmographic models, $(\Omega_\Lambda, h, w_0, w_a)$ in $\wcdm$, and $(\Omega_\Lambda, h)$ in $\axion$]. Since the best fit point fits the data quite well in each of the six tests, one can also use the measurement to represent the best fit theory point for simplicity, 
%
\begin{align}
  \label{eq:best-fit-dist-modulus}
    \mu_{\rm bf}
  & = \frac{5}{2(\gamma - 1)}
    \left [ \log \left (\frac{f_{\rm X}}{f_{\rm UV}^\gamma} \right )
    - \beta
    \right ]
    \cr
  & ~~~~
    - 5\, \log[(4\pi)^{1/2}]
    - 5\, \log \left [{10\,\mathrm{pc}}/{\mathrm{cm}} \right ],
\end{align}
%
where $f_{\rm X,UV}$ are the X-ray and UV band fluxes normalized in units of $\mathrm{erg/s/Hz/cm^2}$.\footnote{In the case of the axion model, they are corrected for extra attenuation due to the axion-photon conversion: $f_{\rm X,UV} = \frac{F_{\rm X,UV}(z)/(\mathrm{erg/s/Hz/cm^2})}{P_{\gamma \gamma \rm ~ X,UV}(z) }$.} 
The error of $\mu_{\rm bf}$ is computed as follows.
\begin{align}
  \label{eq:error-of-mu}
  \Delta & \mu_{\rm bf}
   = \bigg [
    \left ( \frac{\partial \mu}{\partial \log F_{\rm X}}\right )^2 \Delta \log F_{\rm X}^2
    + \left ( \frac{\partial \mu}{\partial \log F_{\rm UV}}
    \right )^2 \Delta \log F_{\rm UV}^2
    \cr
  &  + \left ( \frac{\partial \mu}{\partial \log m_a}\right )^2 \Delta \log(m_a)^2
    +
    \left ( \frac{\partial \mu}{\partial \log g_{a\gamma}}\right )^2 \Delta \log(g_{a\gamma})^2
    \cr
&     + \left ( \frac{\partial \mu}{\partial \beta_0} \right )^2 \Delta \beta_0^2
  + \left ( \frac{\partial \mu}{\partial \gamma} \right )^2
  \Delta \gamma^2
  + \left ( \frac{\partial \mu}{\partial \delta} \right )^2
  \Delta \delta^2
  \bigg ]^{1/2}, 
\end{align}
where $\delta$ is fitted as a nuisance parameter to account for the intrinsic scattering in the quasar dataset. See Appendix~\ref{sec:quasar-data-set} for the explicit likelihood function.
%
We define a distance between the best fit ($\mu_{\rm bf}$) and the concordance cosmology ($\mu_{\rm c}$) as
\begin{align}\label{eq:chi2c}
 \chi^2_{\rm c}
  & = \sum _i \left ( \frac{\mu_{\rm bf, i} - \mu_{\rm c, i}}{\Delta_i} \right )^2,
\end{align}
where $i$ is the index of each QSO data point, and $\Delta = \sqrt{\Delta \mu_{\rm bf}^2 + \Delta \mu_{\rm c}^2} \approx \Delta \mu_{\rm bf}$, since $\Delta \mu_{\rm bf} \gg \Delta \mu_{\rm c}$.
In practice, we neglect the contribution to $\Delta \mu_{\rm bf}$ from $\Delta \log(m_a)$  and  $\Delta \log(g_{a\gamma})$ for simplicity. Since the purpose of $\chi_c^2$ is to quantify the compatibility between the best fit point and the concordance cosmology, neglecting both makes the estimate conservative, \textit{i.e.} it is easier to spot any potential incompatibility.

We show the theory distance as defined above in the last column of Tab.~\ref{tab:bcdm-bsm-fits}. 
Larger value of $ \chi^2_c$ show a larger incompatibility between the best fit point and the concordance cosmology expansion histories. In $w$CDM and the cosmographic model, the incompatibility is much larger than the axion model even after adding SNIa and BAO. 
This is because the axion model mitigates some of the incompatibility by correcting the quasar flux evolution with the axion-induced attenuation $P_{\gamma\gamma,\rm UV,X}(z)$. After this correction the inferred expansion history becomes similar to the concordance cosmology. By contrast, in the other two models the quasar data bends the luminosity distance $D_L(z)$ away from the concordance $\lcdm$ directly, leading to a larger incompatibility between the best fit points and the concordance cosmology.

\subsection{Comment on the Axion Model}
\label{sec:more-axion-model}
\begin{figure}[ht]
  \centering
  \includegraphics[width=.4\textwidth]{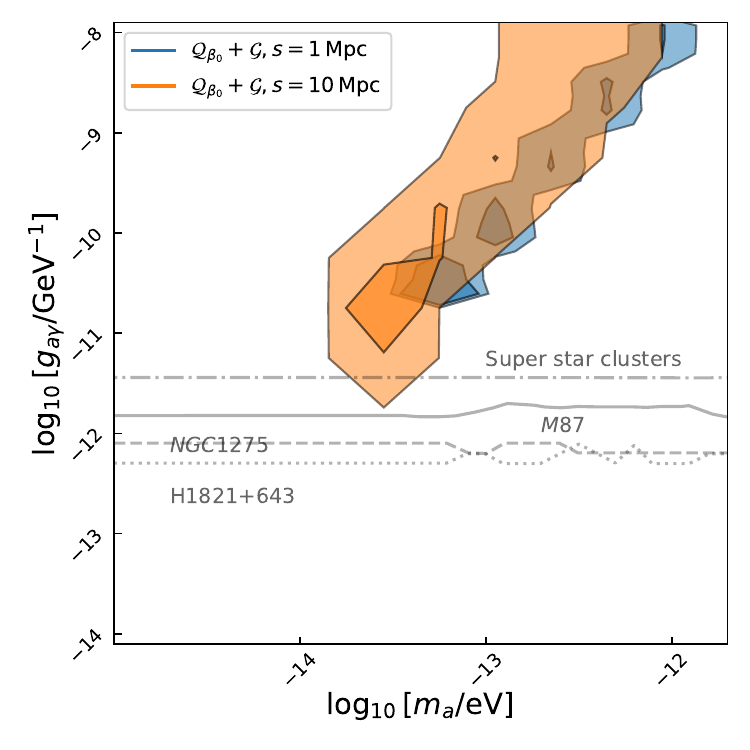}
  \caption{2D posteriors of the axion mass ($m_a$) and coupling to photons ($g_{a \gamma}$) in the $\axion(\Q+\G)$ model, with a magnetic domain size fixed at a comoving length of $s=1~\mathrm{Mpc}$ (blue) and $s=10~\mathrm{Mpc}$ (orange). The darker (lighter) region corresponds to the $1\,\sigma$ ($2\,\sigma$) confidence of the posterior. The gray curves correspond to other astrophysical constraints, the super star clusters~\cite{Dessert:2020lil}, M87~\cite{Marsh:2017yvc}, NGC~1275\cite{Reynolds:2019uqt}, and H1821+643~\cite{Reynes:2021bpe}. }
  \label{fig:axion-bounds}
\end{figure}

As demonstrated in the last section, the axion model can alleviate the tension between the quasar and other cosmological datasets, and its best fit point is more consistent with the concordance cosmology. This motivates us to examine further the posterior of the axion model fit. 

In Fig.~\ref{fig:axion-bounds}, we compare the 2D posterior of the axion parameters with other independent astrophysical constraints~\cite{Reynes:2021bpe,Reynolds:2019uqt,Marsh:2017yvc,Dessert:2020lil}. We observe that such large coupling is disfavored by those constraints. See Fig.~\ref{fig:corner-plots} for more detailed 2D posterior information, and Tab.~\ref{tab:blcdm-bsm-best-fit-appendix} for the full 1D posterior. 

On the other hand, since these constraints rely on galactic magnetic field or the magnetic field in the intracluster medium, the uncertainty in the domain size of the IGM magnetic field does not affect these bounds. As we discussed in Sec.~\ref{sec:beyond-sm}, the allowed domain size of the IGM magnetic field can vary from $1~\mathrm{Mpc}$ to $10~\mathrm{Mpc}$. Because of this, we also fit the axion model with a comoving IGM magnetic field domain length set to $s=10~\mathrm{Mpc}$. This results in a more efficient photon-to-axion conversion, as we showed in Ref.~\cite{Buen-Abad:2020zbd}. Therefore, the model requires a smaller $g_{a\gamma}$ to generate the redshift-dependent feature preferred by the QSO dataset. The results from taking $s=10~\mathrm{Mpc}$ are still in $\sim 2\sigma$ tension with H1821+643~\cite{Reynes:2021bpe} and NGC~1275~\cite{Reynolds:2019uqt},\footnote{The intracluster medium magnetic field modeling for NGC~1275 bound is questioned in~\cite{Libanov:2019fzq}.} although they are still allowed by M87~\cite{Marsh:2017yvc} and Super Star Cluster~\cite{Dessert:2020lil}. Another strong constraint on the axion-photon coupling comes from the CMB spectral distortion~\cite{Mirizzi:2009nq,Tashiro:2013yea}. While a large swath of axion mass values is thereby ruled out, the region $10^{-14}\,\mathrm{eV} \lesssim m_a \lesssim 10^{-12}\,\mathrm{eV}$ can be allowed if {\it (i)} multiple resonant conversions, and {\it (ii)} tuning of the single resonant conversion probabilities both take place \cite{Tashiro:2013yea}.

\begin{figure}[ht]
  \centering
  \includegraphics[clip, trim=0.3cm 1.5cm 0.2cm 2.5cm, width=.49\textwidth]{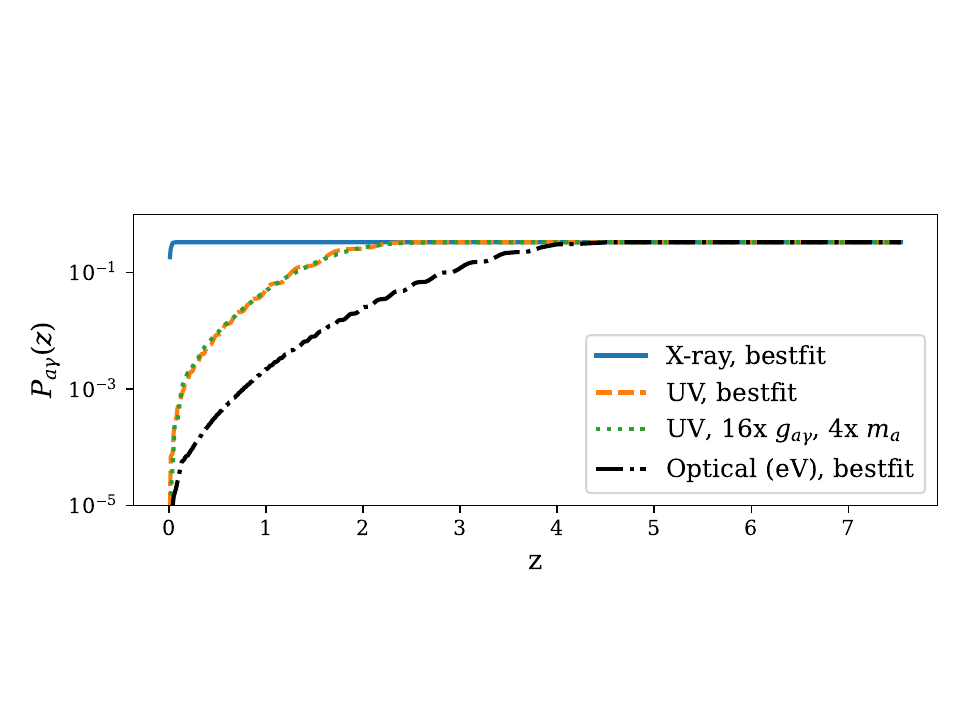}
  \caption{Photon-to-axion conversion probability, at redshifts corresponding to all quasars, in the X-ray band ($\omega = 2~\mathrm{keV}$, blue), UV band ($\omega = 4.96~\mathrm{eV}$, orange), and optical band ($\omega = 1\,\mathrm{eV}$, black), for the axion($\Q+\G$) best fit point. We also recompute the UV photon survival probability by increasing $g_{a\gamma}$ by 16 times and $m_a$ by 4 times in the orange curve. This shows that $P_{\rm UV}(z)$ is almost invariant along the curve $(g_{a\gamma}/g_{a\gamma,\rm bf}) =(m_a/m_{a,\rm bf})^2$, where $(\cdot )_{\rm bf}$ is the best fit point of the parameter. The domain size is chosen to be $s=1~\mathrm{Mpc}$. 
  }
  \label{fig:photon-survival-prob}
\end{figure}

As an extra test to understand the axion-induced photon attenuation, we plot out the photon-to-axion conversion probability at the best fit point of axion($\Q+\G$) in Fig.~\ref{fig:photon-survival-prob}, for different frequencies.
The conversion is very effective in the X-ray frequency range, which transitions from zero to 1/3, the saturated value, mostly below $z\sim0.3$. The conversion in the UV, on the other hand, is almost negligible below $z\sim 1$, merely at the (sub)percent level, while it has a large redshift dependence at $1\lesssim z \lesssim 2.5$. Above $z\sim 2.5$ the UV conversion also saturates at 1/3. Put together, the change in $F_{\rm X}$ and $F_{\rm UV}$ leads to a modification of Eq.~\eqref{eq:1}, i.e. the UV--X-ray flux relation, with Eq.~\eqref{eq:beta-eff-axion}. Such a modification is favored by the QSO data due to its tendency to prefer what amounts to an effective change in $\beta$ around redshift $z \sim 2$, as shown by the $\lcdm(\Qs+\G)$ and $\lcdm(\Qs+\B+\G)$ tests in Sec.~\ref{sec:an-effect-evol}; we discuss this point in more detail in the next section. We further test that the axion-to-photon conversion in the UV band is in the non-linear minimal mixing regime (\textit{i.e.} $k \approx \Delta_a \gg 1/s$ in Eq.~(\ref{eq:pag}) with $s$ being the magnetic domain size). In this regime, the conversion probability has a power-law dependence on both the axion mass and the coupling, $P_{a\gamma} \propto g_{a\gamma}^2/m_a^4$. To verify this, we shift the axion parameters in the direction of $g_{a\gamma} \propto m_a^2$, away from the best fit point. We show that the axion-to-photon conversion is almost invariant along this curve in the $m_a-g_{a\gamma}$ plane, as shown by the green and orange curves in Fig.~\ref{fig:photon-survival-prob}. Note that this behavior corresponds to the degeneracy exhibited by the 2D posterior contours in the $m_a-g_{a \gamma}$ parameter space of Fig.~\ref{fig:axion-bounds}.

We can also understand why the axion island is allowed even after SNIa is added. In Fig.~\ref{fig:photon-survival-prob} we show the impact of the photon-to-axion conversion in the optical band (black points). Below $z\sim 2$, which is the maximum redshift to which the Pantheon dataset extends, the modification of the photon flux in optical band is minimal, at most at the $\sim 2\%$ level. Therefore, this modification hardly generates any perceivable changes in the SNIa dataset even though the coupling is relatively large. 

Lastly, we want to comment on a variation of the axion model, which might evade the astrophysical constraints listed in Fig.~\ref{fig:axion-bounds}. It is also possible to have axion-photon conversion in a cosmic {\it dark} magnetic field $B^\prime$, in the scenario with a dark photon, $\gamma^\prime$, and an axion-photon-dark photon coupling $g_{a \gamma^\prime} a F \tilde{F^\prime}$, where $\tilde{F^\prime}$ is the dual field strength of the gauged dark $U(1)^\prime$. 
 In this case, the conversion probability in a single magnetic domain is similar to Eq.~\eqref{eq:pag} and \eqref{eq:oscs} with $g_{a\gamma} B$ replaced by $g_{a \gamma^\prime} B^\prime$. Assuming that the coherent length of $B^\prime$ is similar to that of the IGM magnetic field ($\sim {\cal O}(1- 10)$ Mpc), from Fig.~\ref{fig:axion-bounds}, we need $g_{a\gamma^\prime} B^\prime \gtrsim 10^{-12}$ GeV$^{-1} \times 1$ nG to restore the cosmic concordance. The astrophysical constraints on $g_{a\gamma^\prime}$ and $B^\prime$ are weaker compared to those on $g_{a\gamma}$ and the IGM $B$ field. More specifically, $B^\prime$ could be as large as micro-Gauss, with the constraints mostly from $N_{\rm eff}$ while $g_{a \gamma^\prime} \lesssim 5 \times 10^{-10}$ GeV$^{-1}$ due to star cooling~\cite{Choi:2018mvk,Choi:2019jwx}. Note that stronger bounds were derived using neutrino magnetic dipole moment, $g_{a\gamma'\gamma} < 1.75 \times 10^{-10}\,\mathrm{GeV}^{-1}$~\cite{Kalashev:2018bra,Capozzi:2020cbu,Carenza:2023qxh}.  This implies a potential new physics explanation for the cosmic concordance of the quasar data, without conflicts with the other constraints. We leave this for further investigation in future work.




\subsection{Effective Evolution of $\beta$}
\label{sec:effect-evol-beta}

\begin{figure}[th]
  \centering
  \includegraphics[clip, trim=0.3cm 1.2cm 0.2cm 1.8cm, width=.49\textwidth]{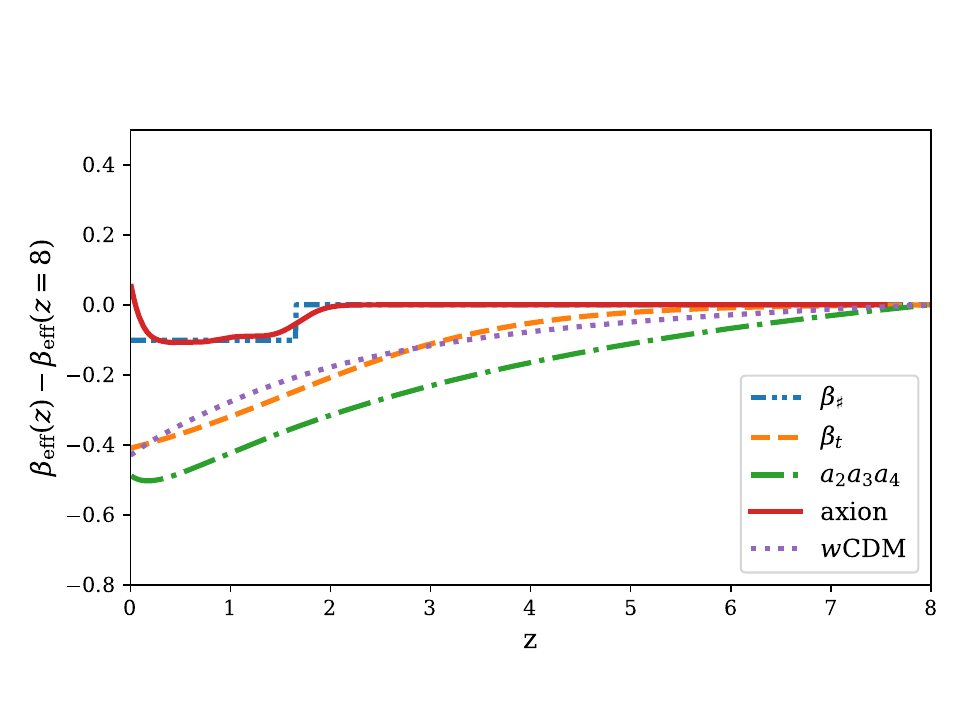}
  \includegraphics[clip, trim=0.3cm 1.2cm 0.2cm 1.8cm, width=.49\textwidth]{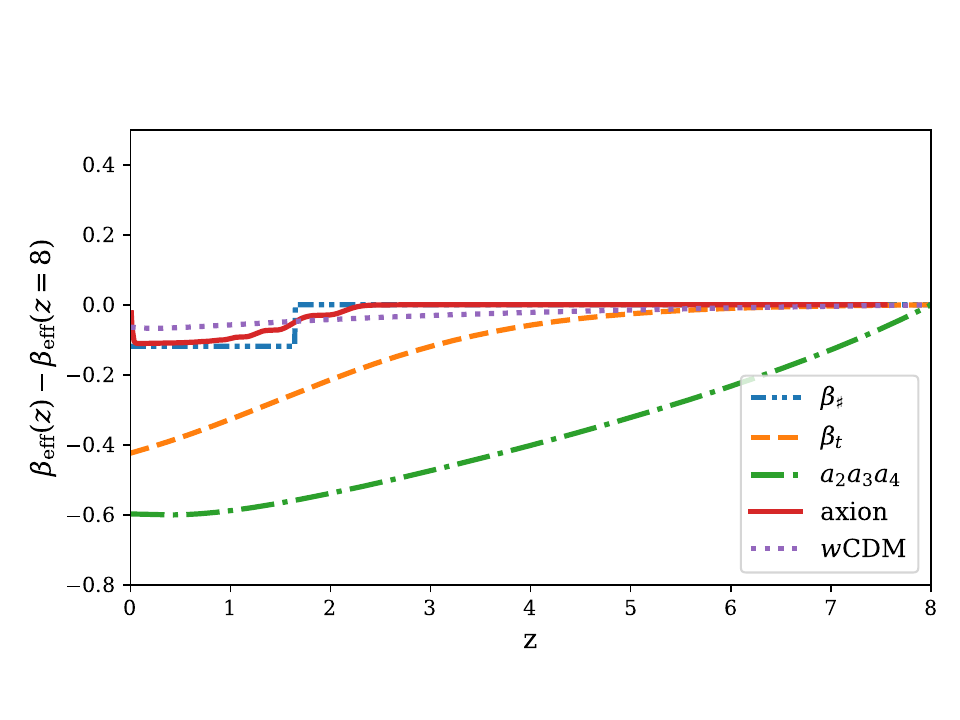}  
  \caption{
    Effective redshift dependence of the nuisance parameter $\beta_\eff(z)$. We show the amount of change in $\beta_\eff(z)$ by measuring it with respect to its value at large $z$ (in this plot, $z=8$). In the upper plot we show this difference for the best fit points of the runs for $\lcdm(\Qs+\G)$, $\lcdm(\Qz+\G)$, $\cg(\Q+\G)$, axion$(\Q+\G)$, and $\wcdm(\Q+\G)$. In the lower panel we do the same but including SNIa and BAO in the fits ($+\B$ dataset).
    We find a good agreement between our results that make use of $\Qz$, and the effective evolution $\beta_{\rm eff}$ of Eq.~(\ref{eq:beta-eff-from-ep-correction}) with parameter values taken from Ref.~\cite{Dainotti:2022rfz}. 
  }
  \label{fig:beta-evo}
\end{figure}

As we discuss in Sec.~\ref{sec:how-NP-bias-quasars}, new physics modifications can be parametrized as an effective redshift evolution in $\beta$. We summarize the results by showing this effective redshift evolution, $\beta_{\rm eff}(z)$, in Fig.~\ref{fig:beta-evo}.

In Sec.~\ref{sec:an-effect-evol}, when fitted to the QSO data alone, we see roughly two classes of solutions: a shallow but abrupt change in $\beta$ around $z\sim 1.5$ and a deep and slow change of $\beta$ spanning the whole redshift range. The axion model leads to a relatively quick change in redshift through $P_{\gamma\gamma,\rm UV,X}(z)$. This serves as a good physical realization of the sharp change given by the effective parametrization in $\Qs$. As we see in Sec.~\ref{sec:more-axion-model}, the effective change in $\beta$ around $z= 1.5$ is due to the UV photon conversion to axions. Interestingly, the maximum amount of change allowed in the UV is given by
\begin{align}
  \label{eq:axion-change-beta-maximal}
  \Delta \beta  = \gamma \log (2/3) \approx -0.12. 
\end{align}
This size of the modification in the UV-X-ray relation happens to be what the data prefers, as evidenced by the run with the effective parametrization with a sharp transition in $\beta$ ($\beta_\sharp$ in the plot). 
We would like to highlight the fact that when the QSO dataset is allowed to have a sharp transition in the UV-X-ray relation (\textit{i.e.} $\Qs$), the fits prefer this transition to take place at around the same redshift as our axion model does.
The striking agreement between both can be seen by comparing $\lcdm(\Qs+\G)$ with $\axion(\Q+\G)$, or by comparing $\lcdm(\Qs+\B+\G)$ and $\axion(\Q+\B+\G)$.
Adding to its appeal, the axion model outperforms non-standard cosmological models, even those as flexible as the cosmographic model, as we show in Sec.~\ref{sec:results-blcdm-bsm}.





By contrast, the $w$CDM and cosmographic models have large changes when $\B$ set is added. One should keep in mind that in the $\blcdm$ models, the effective evolution in $\beta$ is achieved at the price of directly modifying the Universe expansion history. Therefore, SNIa+BAO will constrain the low-$z$ part ($z\lesssim 1.5$) of $\beta_{\rm eff}(z)$ to be as flat as possible. This has consequences in both the $w$CDM and the cosmographic model. In $w$CDM, having a flat low-$z$ part of $\beta_{\rm eff}(z)$ restricts the overall amount of change one can achieve from $z=0$ to $z\sim 8$. This is reflected by the change in the curve for $\wcdm(\Q+\G)$ and $\wcdm(\Q+\G+\B)$: adding $\B$ significantly reduces the amount of change in $\beta_{\eff}(z)$.
On the other hand, the cosmographic model has enough degrees of freedom to guarantee a relatively flat low-$z$ $\beta_{\rm eff}(z)$. Therefore, while restricting the flatness at low-$z$ changes the shape of the effective $\beta$ evolution, the large modification in $\beta$ survives. However, this large $\beta$ change itself comes at the price of a large deviation from the concordance $\lcdm$ at $1 \lesssim z \lesssim 8$. It also remains unclear what physical models can achieve such a modification in $D_L(z)$ as indicated by the cosmographic model.

\section{Conclusion}
\label{sec:conclusion}

  Although quasar data can be used to extract information about the late-time cosmological history of the Universe, it stands in need of standardization. In light of the recent efforts along these lines, we perform a critical examination on the imprints that new physics may leave on the quasar data, in the form of unaccounted-for redshift evolution of their fluxes. This is a timely endeavor, given the ongoing Hubble tension, as well as the discrepancy between the cosmological expansion inferred from quasars and other observations.

We use a flexible fitting template to identify what is required to modify the quasar flux in such a way that the concordance cosmological history given by in Eq.~(\ref{eq:concordance}) can be restored. We test $\wcdm$ and the cosmographic model, two cosmological models beyond $\lcdm$, as well as the axion model, a mechanism beyond the SM that can present extra photon attenuation. We find that the axion model has an advantage over the other two, restoring the consistency between the best-fit parameters emerging from the quasar dataset, and the concordance cosmology. It also outperforms the $\blcdm$ models in improving the goodness of fit to the quasar dataset alone.

{We stress that, while the axion model is preferred over the other non-standard cosmological models, the best-fit point is in tension with several astrophysical constraints of the axion-photon coupling. As a further test, we have taken into account the uncertainty in the magnetic field in the IGM and varied its coherent domain size. We see that the tension between the quasar-preferred axion theory point and current bounds is somewhat relieved but not completely resolved. We comment in passing that it remains possible that the axion-photon conversion takes place in a {\it dark} magnetic field, which may relax some of the astrophysical bounds which our results are in tension with.} These variants of our baseline axion model may evade other constraints while at the same time providing the redshift evolution of the quasar fluxes required by observation, maintaining the validity of the RL relation, and restoring the consistency of the quasar data with the concordance cosmological model.


\textit{Acknowledgment~~~~}We thank Daniele Alves, Michael Graesser, Hui Li, Nicole Lloyd-Ronning, and Elisabeta Lusso, for useful comments. We also thank Edoardo Vitagliano, Eoin \'O Colg\'ain, and Bharat Ratra for useful inputs on an early version of the draft.
We thank Tomer Volansky for the access to Tel Aviv Univeristy High-Performance Computing, {and the Center for Computation \& Visualization at Brown University} where part of the work was performed.
This work was supported by the U.S. Department of Energy through the Los Alamos National Laboratory. Los Alamos National Laboratory is operated by Triad National Security, LLC, for the National Nuclear Security Administration of U.S. Department of Energy (Contract No. 89233218CNA000001). Research presented in this article was supported by the Laboratory Directed Research and Development (LDRD) program of Los Alamos National Laboratory under projects 20220135DR and 20230047DR.
MBA is supported in part by the National Science Foundation under Grant Number PHY-2210361, and the Maryland Center for Fundamental Physics. JF is supported by the NASA grant 80NSSC22K081 and the DOE grant DE-SC-0010010.

\appendix

\begin{figure}[htp]
  \centering
  \includegraphics[width=.42\textwidth]{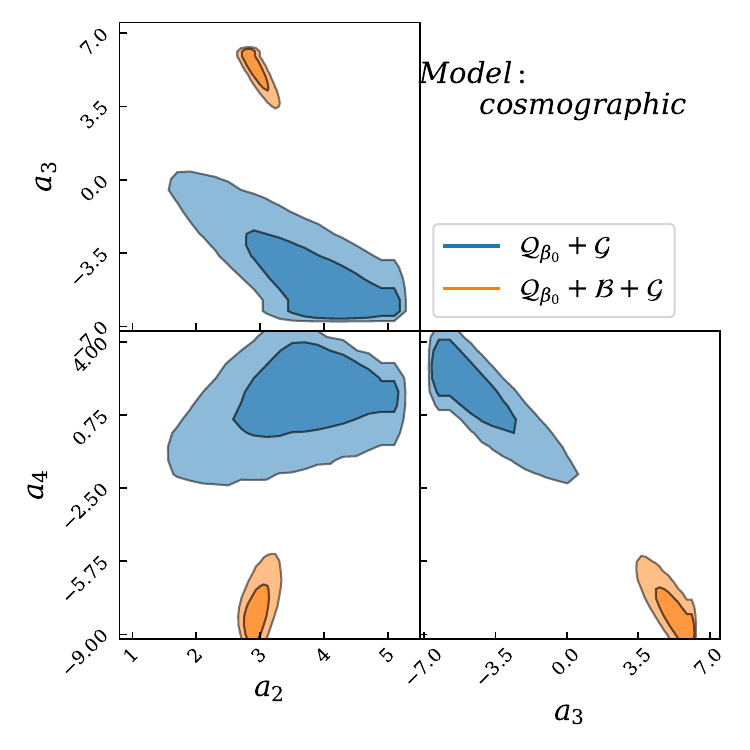}\\
  \vspace{-0.25cm}
  \includegraphics[width=.42\textwidth]{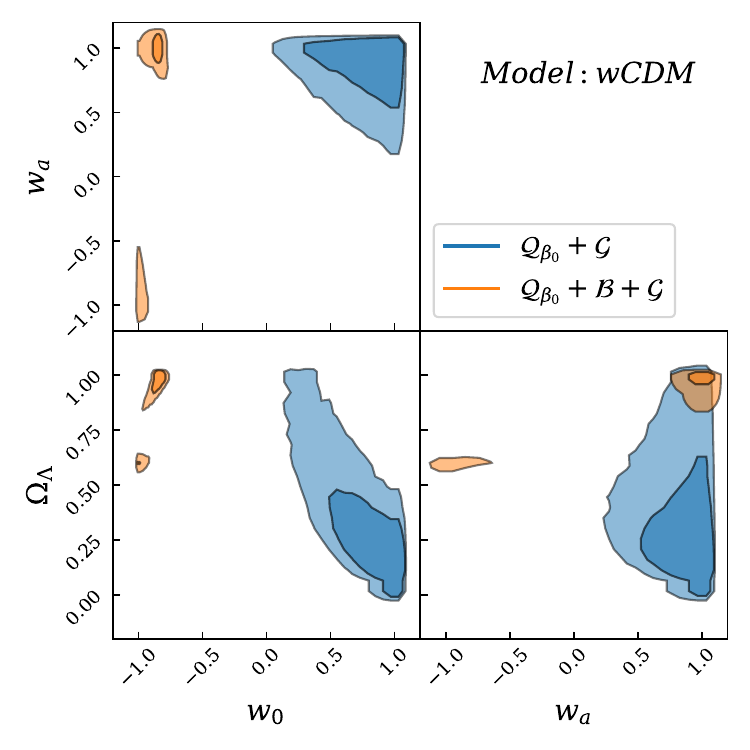}\\
  \vspace{-0.25cm}  
  \includegraphics[width=.42\textwidth]{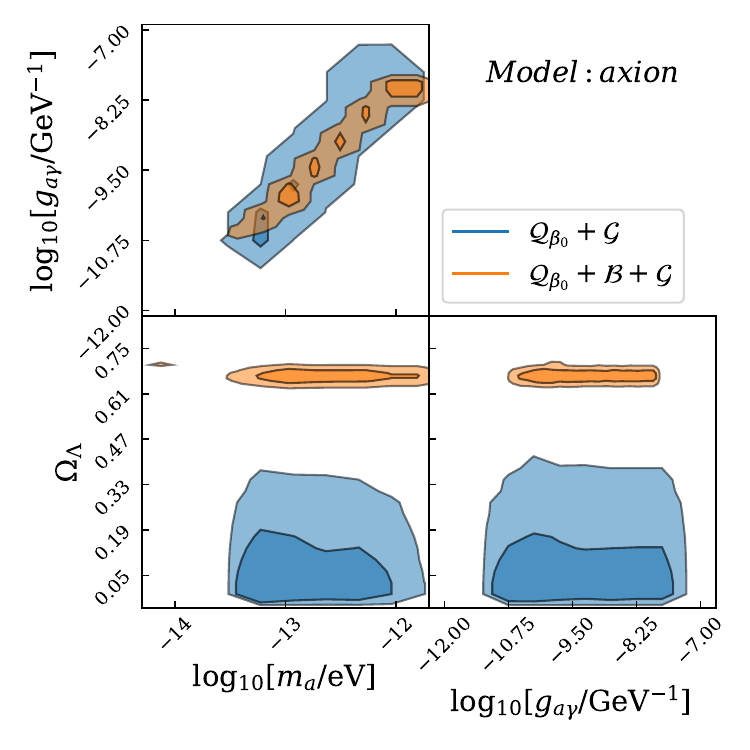}
  \vspace{-0.25cm}    
  \caption{
    2D posteriors of the parameters of the cosmographic (top panel), $\wcdm$ (middle panel), and axion (bottom panel) models. The magnetic domain has a comoving size $s=1~\mathrm{Mpc}$. {The darker (lighter) region corresponds to the $1\,\sigma$ ($2\,\sigma$) confidence in the posterior.} 
  }
  \label{fig:corner-plots}
\end{figure}

\section{Quasar Dataset}
\label{sec:quasar-data-set}

We use the quasar data sample from Ref.~\cite{Lusso:2020pdb}. It contains seven different groups: the XMM-Newton $z \sim 3$ sample, the new XMM-Newton $z \sim 4$ quasars, the High-$z$ sample, XXL, SDSS - 4XMM, SDSS - Chandra, and local AGN.

The UV--X-ray relation is expected to hold up to a certain amount of scattering that cannot be eliminated by the improvement of the measurement. Since this intrinsic scattering of the quasars is unknown, we add a third nuisance parameter, $\delta$, describing the standard deviation of the intrinsic scattering, along with $\gamma$ and $\beta$. As a result, the log-likelihood function is given by
\begin{align}
  \label{eq:likelihood}
  -2\, \ln(\mathrm{likelihood})
  & =
    \sum_i \left [  \left  (x^{\rm th} - x^{\rm obs}\right )^2/\sigma_i^2  + \ln(2\pi \sigma^2) \right ],
\end{align}
where $\delta$ is combined into the error of each quasar flux measurement as $\sigma_i^2 = \sigma_{i,  \rm \log F_{\rm UV,X}}^2 + \delta^2$.
By adopting Eq.~\eqref{eq:likelihood}, $\delta$ can be treated as a nuisance parameter that gets fitted together in the MCMC. Therefore, specific to the QSO dataset, we have the following nuisance parameters: $\beta,\gamma,\delta$, with $\beta$ itself parametrize in the various ways described in Sec.~\ref{sec:an-effect-evol}.

\section{Methodology}
\label{sec:methodology-1}
We use \texttt{emcee}~\cite{Foreman_Mackey_2013} to fit the three models ($\wcdm$, cosmographic, and $\axion$) to two data combinations ($\Q+\G$ and $\Q+\B+\G$).  Our code is publicly available at \href{https://github.com/ChenSun-Phys/high\_z\_candles.git}{{\tt github.com/ChenSun-Phys/high\_z\_candles}}. We use 100 walkers and a chain length of 40000.
All runs pass our MCMC convergence condition of the chain length to be 50 times longer than the auto-correlation length, except for $\wcdm(\Q+\B+\G)$, where we loosen the condition to 12 times the auto-correlation length due to its strong bimodal posterior.

The results are analyzed with \texttt{corner.py}~\cite{corner}. Since different parameters have different auto-correlation times, we choose the burn-in to be equal to twice the largest one, and choose the thinning-length to be equal to half of the smallest one.

We use the following flat priors for the theory parameters.
\begin{align}
  \label{eq:priors-model}
  & \Omega_\Lambda
   \in (0, 1) ~~~ h \in (0.6, 0.8) \\
  &  w_0
   \in (0, 1) ~~~ w_a \in (0, 1) \cr
  &  a_2
   \in (1, 5) ~~ a_3 \in (-6, 6) ~~ a_4 \in (-10, 6)
    \cr
  &
    \log \left  [\frac{g_{a\gamma}}{\mathrm{GeV}^{-1}} \right ] \in (-18, -8) ~~
    \log \left  [\frac{m_a}{\mathrm{eV}} \right ] \in (-17, -11).
    \notag
\end{align}
We set the following priors for the nuisance parameters when the corresponding datasets are used.
\begin{align}
  \label{eq:priors-nuisance}
  & \mathrm{SNIa}: M_0 \in (-21, -18)
  \\
  & \mathrm{BAO}: r_s/\mathrm{Mpc} \in (120, 160)
  \cr
  & \mathrm{QSO}:  \gamma \in (0.1, 1) ~~ \delta \in (0.05, 0.6)
    \cr
  & ~ \Q: \beta_0 \in (0, 10)
    \cr
  & ~ \Qs:  \beta_0 \in (0, 10) ~~ \beta_1 \in (0, 10) ~~ z_0 \in (0, 9)
    \cr
  & ~ \Qz: \beta_0 \in (0, 10) ~~ \beta_1 \in (0, 10) ~~ z_0 \in (0, 9) ~~ \delta z \in (0.01, 10)~.
    \notag
\end{align}

\begin{table*}[t]
  \centering
  \begin{tabular}{c|c c c c c c c c c c }
    \hline 
    & $\Omega_\Lambda$ & $h$ & $\beta_0$ & $\beta_1$ & $z_0$ & $\delta z$ & $\gamma$ & $\delta$ & $M_0$ & $r_s \mathrm{[Mpc]}$ \\
    \hline
    $\B+\G$ &  $0.68^{+0.02}_{-0.02}$ & $0.676^{+0.004}_{-0.004}$  &  & & & & & & $-19.41^{+0.01}_{-0.01}$ & $147.12^{+0.26}_{-0.26}$\\
    $\Q+\G$   & $0.05^{+0.07}_{-0.03}$ & $0.674^{+0.005}_{-0.005}$& $7.04^{+0.24}_{-0.24}$ & & & & $0.639^{+0.008}_{-0.008}$ & $0.228^{+0.004}_{-0.003}$ & &  \\
    $\Qs+\G$  &  $0.10^{+0.12}_{-0.07}$ & $0.674^{+0.005}_{-0.005} $ & $8.20^{+0.28}_{-0.28}$ & $8.30^{+0.29}_{-0.29}$ & $1.65^{+0.01}_{-0.01} $& &  $0.599 ^{+0.009}_{-0.009} $ & $0.225^{+0.003}_{-0.003}$ & \\
    $\Qz+\G$   &  $0.46^{+0.26}_{-0.27}$ & $0.674^{+0.005}_{-0.005}$ & $8.38^{+0.37}_{-0.36}$ & $9.37^{+0.35}_{-0.36}$ & $4.91^{+3.27}_{-2.05}$ & $1.24^{+1.13}_{-0.86}$ &  $0.580^{+0.011}_{-0.010}$ & $0.224^{+0.003}_{-0.003}$\\
    $\Q+\B+\G$ & $0.66 ^{+0.02}_{-0.02}$ & $0.674 ^{+0.004}_{-0.004}$ & $6.40^{+0.23}_{-0.23}$ &&&&$0.662^{+0.008}_{-0.008}$ & $0.230^{+0.004}_{-0.004}$ & $-19.41^{+0.01}_{-0.01} $ & $147.09^{+0.26}_{-0.26}$ \\
    $\Qs+\B+\G$ & $0.67^{+0.02}_{-0.02}$ & $0.675^{+0.004}_{-0.004}$ & $7.79^{+0.27}_{-0.27}$ & $7.90^{+0.28}_{-0.28}$ & $1.65^{+0.01}_{-0.01}$ & & $0.616^{+0.009}_{-0.009}$ & $0.226^{+0.003}_{-0.003}$ & $-19.41^{+0.01}_{-0.01}$ & $147.10^{+0.26}_{-0.26}$ \\
    $\Qz+\B+\G$ & $0.68^{+0.02}_{-0.02}$ & $0.676^{+0.004}_{-0.004}$ & $8.40^{+0.39}_{-0.40}$ & $9.40^{+0.38}_{-0.37}$ & $4.30^{+2.75}_{-1.52}$ & $0.92^{+0.76}_{-0.64}$ & $0.579^{+0.011}_{-0.011}$ & $0.224^{+0.003}_{-0.003}$ & $-19.41^{+0.01}_{-0.01}$ & $147.12^{+0.26}_{-0.26}$\\
    \hline 
  \end{tabular}
  \caption{\label{tab:lcdm-best-fit-appendix}
  The mean and $1\sigma$ values of the $\lcdm$ parameters, fitted to various data combinations.}
\end{table*}

\begin{table*}[t]
  \centering
  \begin{tabular}{c|c c c c c c c c c c c c c c}
    \hline 
    cosmographic & $h$ & $a_2$ & $a_3$ & $a_4$   &  $\beta_0$ & $\gamma$ & $\delta$ & $M_0$ & $r_s \mathrm{[Mpc]}$ \\
    \hline
    $\Q+\G$    &  $0.674^{+0.005}_{-0.005}$&  $3.66^{+0.71}_{-0.86}$& $-4.35^{+1.97}_{-1.19}$ & $1.61^{+1.25}_{-1.69}$  & $8.81^{+0.32}_{-0.32}$ & $0.579^{+0.011}_{-0.011}$ & $0.224^{+0.004}_{-0.003}$   \\
    $\Q+\B+\G$  &  $0.677^{+0.005}_{-0.005}$& $2.95^{+0.15}_{-0.12}$ & $5.19^{+0.55}_{-0.78}$ & $-8.01^{+1.05}_{-0.82}$  & $7.67^{+0.27}_{-0.27}$ & $0.619^{+0.009}_{-0.009}$ & $0.226^{+0.003}_{-0.003}$  & $-19.42^{+0.01}_{-0.01}$ & $147.13^{+0.26}_{-0.26}$\\
    \hline 
  \end{tabular}
  ~\\~\\~\\~\\
  \begin{tabular}{c|c c c c c c c c c c c c c c}
    \hline 
    $\wcdm$ & $\Omega_\Lambda$ & $h$  & $w_0$ & $w_a$ &  $\beta_0$ & $\gamma$ & $\delta$ & $M_0$ & $r_s \mathrm{[Mpc]}$ \\
    \hline
    $\Q+\G$   & $0.28^{+0.28}_{-0.13}$ & $0.674^{+0.005}_{-0.005}$ &  $0.77^{+0.17}_{-0.29}$ & $0.83^{+0.12}_{-0.24}$ & $8.61^{+0.30}_{-0.31}$ & $0.58^{+0.01}_{-0.01}$ & $0.225^{+0.003}_{-0.003}$ \\
    $\Q+\B+\G$ & $0.96^{+0.02}_{-0.08}$ & $0.675^{+0.005}_{-0.004} $ & $-0.86^{+0.03}_{-0.06}$ & $0.96^{+0.03}_{-0.10}$ & $6.68^{+0.25}_{-0.25}$ & $0.653^{+0.008}_{-0.008}$ & $0.229^{+0.004}_{-0.003}$ & $-19.43^{+0.015}_{-0.014}$ & $147.10^{+0.26}_{-0.25}$ \\
    \hline 
  \end{tabular}
  ~\\~\\~\\~\\
  \begin{tabular}{c|c c c c c c c c c c c c c c}
    \hline 
    axion & $\Omega_\Lambda$ & $h$  &  $\log(m_a)$ & $\log(g_{a\gamma})$ &  $\beta_0$ & $\gamma$ & $\delta$ & $M_0$ & $r_s \mathrm{[Mpc]}$ \\
    \hline
    $\Q+\G$   & $0.09^{+0.12}_{-0.06}$ & $0.674^{+0.005}_{-0.005}$ & $-12.85^{+0.58}_{-0.46}$ & $-9.77^{+1.21}_{-0.80}$& $8.18^{+0.25}_{-0.28}$ & $0.606^{+0.009}_{-0.008}$ & $0.225^{+0.004}_{-0.003}$ \\
    $\Q+\B+\G$ & $0.67^{+0.02}_{-0.02}$ & $0.675^{+0.005}_{-0.004}$ & $-12.71^{+0.47}_{-0.37}$ & $-9.39 ^{+0.95}_{-0.77}$ & $7.68^{+0.23}_{-0.22}$ & $0.625^{+0.007}_{-0.007}$ & $0.226^{+0.004}_{-0.003}$ & $-19.41^{+0.01}_{-0.01}$ & $147.11^{+0.26}_{-0.26}$\\
    \hline 
  \end{tabular}  
  \caption{\label{tab:blcdm-bsm-best-fit-appendix}
    The mean and $1\sigma$ values of the parameters of the cosmographic (top), $\wcdm$ (middle), and axion (bottom) models, fitted to various data combinations.}
\end{table*}

\section{Posterior of the Fits}
\label{sec:corner-plots}
We show the 1D, $1\sigma$ posterior range of the $\lcdm$ parameters in Tab.~\ref{tab:lcdm-best-fit-appendix}. 
The corresponding results for the $\blcdm$ and BSM fits are shown in Tab.~\ref{tab:blcdm-bsm-best-fit-appendix}.
The 2D posterior of the $\blcdm$ and BSM parameters are shown in Fig.~\ref{fig:corner-plots}.








\vfill
\pagebreak
~\\~\\
\pagebreak

\bibliography{bib}
\bibliographystyle{utphys}
\end{document}